# Observation of Ising spin-nematic order and its close relationship to the superconductivity in FeSe single crystals

Dongna Yuan,[1] Jie Yuan,[1] Yulong Huang,[1] Shunli Ni,[1] Zhongpei Feng,[1] Huaxue Zhou,[1] Yiyuan Mao,[1] Kui Jin,[1,2] Guangming Zhang,[3] Xiaoli Dong,[1,2,*] Fang Zhou,[1,2,§] and Zhongxian Zhao[1,2]

[1] Beijing National Laboratory for Condensed Matter Physics, Institute of Physics, Chinese Academy of Science, Beijing 100190, China
[2] University of Chinese Academy of Sciences, Beijing 100049, China
[3] State Key Laboratory of Low Dimensional Quantum Physics and Department of Physics, Tsinghua University, Beijing 100084, China

[*] dong@iphy.ac.cn; [§] fzhou@iphy.ac.cn

**Abstract**

Superconducting FeSe single crystals of (001) orientation are synthesized via a hydrothermal ion-release route. An Ising spin-nematic order is identified by our systematic measurements of in-plane angular-dependent magnetoresistance (AMR) and static magnetization. The turn-on temperature of anisotropic AMR signifies the Ising spin-nematic ordering temperature $T_{sn}$, below which a two-fold rotational symmetry is observed in the iron plane. A downward curvature appears below $T_{sn}$ in the temperature dependence of static magnetization for the weak in-plane magnetic field as reported previously. Remarkably, we find a universal linear relationship between $T_c$ and $T_{sn}$ among various superconducting samples, indicating that the spin nematicity and the superconductivity in FeSe have a common microscopic origin.

The tetragonal $\beta$-FeSe was reported to show bulk superconducting transition at $T_c \sim 8$ K [1, 2]. It is notable that its $T_c$ can be enhanced to 36.7 K under a high pressure [3-6] and even to $\sim 48$ K via charge injection [7]. Also, a higher $T_c$ than the binary FeSe is always achieved in ions/clusters intercalated iron selenides. For example, the superconductivity with $T_c \sim 42$ K has been realized in $(Li_{1-x}Fe_x)OHFe_{1-y}Se$ [8], where the two dimensionality of the electronic structure of the iron plane is enhanced due to the expansion in the interlayer separation [9, 10]. In contrast, the bulk FeSe displays the maximum interlayer compactness in the iron-based superconductors and thus the lowest $T_c$.

In FeSe superconductors, no antiferromagnetic long-range order was reported to exist in ambient pressure, but the presence of the rotational symmetry breaking in the electronic structure of the iron plane and its implication for superconducting paring have drawn much attention [11-17]. It has been argued that the tetragonal-to-orthorhombic structure transition at $T_s \sim 90$ K is driven by the ferro-orbital ordering with unequal occupancies of the $3d_{xz}/3d_{yz}$ orbitals [18, 19]. However, the structural transition temperature $T_s$ remains nearly the same for various samples showing different $T_c$'s (see below in Fig. 4). Recent neutron scattering measurements [20-22] suggest that the electron pairing for the superconductivity is closely related to the stripe-like ($\pi$, 0) antiferromagnetic (AFM) spin fluctuations and a sharp spin resonance is observed in the superconducting phase. Therefore, the key issue turns out to be that if any peculiar order of the spin origin showing the rotational symmetry breaking exists and how it is related to the superconductivity in FeSe.

Here we report the presence of an Ising spin-nematic order in our FeSe single crystal samples based on the measurements of angular-dependent magnetoresistance (AMR) and static magnetization. The onset temperature $T_{sn}$ of this nematic order strongly depends on the superconducting transition temperature ($T_c$), and spans a wide range from far below to beyond the structural transition temperature ($T_s$). Our results suggest that the spin nematicity is driven by strongly frustrated spins with the ($\pi$, 0) stripe fluctuations predominating in bulk FeSe. Importantly, a universal linear relationship between $T_c$ and $T_{sn}$ is found among various superconducting samples, indicating that the spin nematicity and the superconductivity in FeSe have a common microscopic origin.

In order to identify the electronic correlations in the iron plane crucial for the superconductivity, sizable FeSe crystal samples of (001) plane orientation with different $T_c$'s are essential. Although the samples of the (001) orientation can be obtained by, e.g., vapor transport growth, it is a very time-consuming process. On the other hand, the high-temperature growth by flux-free floating-zone or flux method only produces the samples with (101) orientation. Most recently, by a high-efficient hydrothermal ion release/introduction technique, we have successfully synthesized large FeSe single crystal samples of the (001) orientation. The details of sample preparation have been reported in ref. 23, similar to the ion/cluster exchange growth of large $(Li_{0.84}Fe_{0.16})OHFe_{0.98}Se$ single crystals [10]. Via this hydrothermal process, the superconducting FeSe single crystal can be derived from the insulating $K_{0.8}Fe_{1.6}Se_2$ matrix. Namely, the interlayer K ions in the matrix are completely *released* and the $\sqrt{5}\times\sqrt{5}$ ordered vacant Fe sites ~ 20% in amount in the $Fe_{0.8}Se$-layers are occupied by *introduced* Fe ions. The end FeSe single crystal naturally inherits the original (001) orientation of the matrix, in which no trace of K is detected by EDX. Powder XRD confirms the pure tetragonal β-FeSe phase with the lattice parameters $a$ = 3.7725(1) Å and $c$ = 5.5247(2) Å for the sample of $T_c$ ~ 7.6 K [23]. The magnetic measurements were conducted on a Quantum Design MPMS-XL1 system of a small remnant field ＜4 mOe. The electrical resistivity and the angular-dependent magnetoresistance were measured on a Quantum Design PPMS-9.

For a typical hydrothermal crystal sample displaying the (001) orientation, the bulk superconductivity at $T_c$ ~ 7.6 K is confirmed by the magnetic and electrical resistivity measurements, shown in Fig. 1 (a) and (b). The high superconducting quality is demonstrated by the sharp diamagnetic transitions as well as the 100% diamagnetic shielding, though the sample shows a crystal mosaic of approximately 5 degrees in terms of the full-width-at-half-maximum of x-ray rocking curve [23]. In this work, we also performed the similar measurements on a flux-grown FeSe crystal sample of (101) orientation exhibiting a higher $T_c$ [24]. Its superconductivity is shown in Fig. 1 (c) and (d). The temperature dependences of the normal state resistivity in the whole measuring temperature range for the two typical samples are displayed in Fig. 1 (e) and (f), respectively. All the $T_c$ values here are determined by the onset temperatures of the diamagnetic transition, defined as that where the shielding and Meissner signals clearly separate from one another. We find that the $T_c$ value of FeSe is sensitive to the carrier concentrations of electron and hole bands from our Hall measurements (to be reported elsewhere).

With our single crystal samples, the angular-dependent magnetoresistance measurements are performed. We fixed the current direction and varied the angle ($\theta$) between the directions of the external field (*H*) and the current (*I*), with $\theta = 0°$ corresponding to $H \perp I$. Remarkably, for both the FeSe crystal samples with the lower $T_c$ ~7.6 K and the higher $T_c$ ~10.0 K, the AMR in the normal state exhibits a *two-fold* rotational symmetry, which are turned on below $T_{sn}$ ~55 K and ~100 K as shown in Fig. 2 (a) and (b), respectively. The anisotropy in AMR is getting enhanced with decreasing temperature. Such an enhancement in charge scatterings is also manifested in temperature-dependent magnetoresistance (MR) [25]. Our observation of the AMR anisotropy with the two-fold rotational symmetry thus provides a decisive evidence for the presence of a nematic ordering different from the ferro-orbital ordering, which is accompanied with the structural transition occurring at almost the same temperature ($T_s$ ~ 90 K) in samples with different $T_c$'s.

Furthermore, a downward curvature below $T_{sn}$ ~ 55 K for the (001) crystal sample of $T_c$ ~7.6 K has been observed in the static magnetization under an in-plane magnetic field of 0.1 T (Fig. 5a in ref. 23). Such a feature is strongly dependent on the magnitude of the magnetic field: it fades out when the field is lowered to 0.01 T (Fig.

5b in ref. 23). This indicates that the strong quantum spin frustrations predominate in the iron plane. Although the orbital ordering below $T_s$ is of the two-fold rotational symmetry, the obvious downward feature of in-plane static magnetization below the characteristic $T_{sn} \sim 55$ K, which is far below $T_s$, suggests that the two-fold anisotropy identified by our AMR measurements is closely related to the frustrated spins with the anisotropic magnetic fluctuations, rather than the orbital ordering. Therefore, we are led to the conclusion that an Ising-like spin-nematic order emerges below $T_{sn}$. The corresponding order parameter is characterized by

$$\sigma = \langle S_i \cdot S_{i+x} - S_i \cdot S_{i+y} \rangle, \tag{1}$$

where $S_{i+x}$ and $S_{i+y}$ stand for the nearest neighbor spins of the spin $S_i$ on the square lattice, respectively. Actually, such Ising spin nematicity is argued to exist in the strongly frustrated limit of the quantum frustrated spin-1 Heisenberg model with nearest neighbor antiferromagnetic coupling and next nearest neighbor antiferromagnetic coupling [12, 16, 26-29]. Consequently, the appearance of the two-fold anisotropy in AMR for the (001) sample is well explicable by the temperature-dependent anisotropic scattering of the charge carriers caused by the spin-nematic order below $T_{sn}$. For the AMR measurement on the sample of (101) orientation, the iron plane component of the magnetic field takes effect. Considering the $(\pi, 0)$ stripe spin fluctuations reported for FeSe superconductors, we argue that the maxima of the anisotropic AMR points in the crystallographic $a$ direction with antiferromagnetic correlations and the minima in the $b$ direction with ferromagnetic correlations.

Moreover, we have also hydrothermally synthesized another FeSe single crystal of (001) orientation, and its characteristic temperatures are determined as $T_c \sim 6.8$ K and $T_{sn} \sim 37.5$ K, shown in Fig. 3. The difference in $T_c$ value results from the difference in concentrations of electron and hole bands on the basis of our Hall resistance measurements (to be reported elsewhere). When summarizing all the data of our three single crystal samples, we found a remarkable linear relationship between $T_c$ and $T_{sn}$

(the dotted straight line in Fig. 4). The fitting gives rise to an expression

$$T_c = \alpha \cdot T_{sn} + T_{min}, \qquad (2)$$

with $\alpha \sim 0.052$ and $T_{min} \sim 4.8$ K. Moreover, we also collect other three sets of $T_c$ and $T_{sn}$ given by either the onset temperature of MR or the cusp temperature of the in-plane magnetic magnetization on FeSe single crystal samples of the (001) orientation [25, 30, 31]. All the collected $T_c$ and $T_{sn}$ well satisfy this linear relationship as well. Meanwhile, the structural transition temperatures ($T_s$'s) by the x-ray or neutron diffractions on various FeSe samples with different $T_c$'s available from the literature [19, 20, 22, 32-36] are plotted in Fig. 4. In contrast to the $T_s$ remaining nearly the same value of ~90 K, the value of the spin-nematic ordering temperature $T_{sn}$ varies with $T_c$ in a wide range from far below to beyond $T_s$. Therefore, the superconductivity and the spin nematicity are correlated by the stripe AFM spin fluctuations, rather than the structural phase transition or the orbital ordering. Interestingly, this universal linear relationship allows the spin-nematic ordering to coincide with the superconducting transition at $T_{min}/(1-\alpha) \sim 5.1$ K, which is worthy of a further study.

It needs to be emphasized that, for the FeSe samples with $T_c$'s around 9.5 K, both the spin nematicity transition and the ferro-orbital ordering/structure transition happen to occur in the vicinity of ~ 90 K, as shown in Fig. 4. So it is very difficult to distinguish experimentally these different ordering transitions in such samples. However, our specific samples cover the $T_c$ values from 6.8 K to 10.6 K, so that the spin-nematic ordering temperature spans from 37.5 K to 120 K, well separated from the structural transition temperature ~ 90 K. Therefore, our results have disentangled the essential role played by the spin-nematic ordering in the superconducting pairing, a long-standing puzzle in bulk FeSe superconductors.

In conclusion, we have experimentally evidenced the emergence of the spin-nematic ordering below $T_{sn}$ in the normal state of the superconducting FeSe single crystals. The universal linear relationship between $T_c$ and $T_{sn}$ has been found for the first time, which spans a wide temperature range. Our results have shed new light on the mechanism of unconventional superconductivity in FeSe, including its drastic enhancement of the superconducting transition temperature under pressure when the nematicity is suppressed.

Note added: we find that the $T_c$ (~10.6 K) and $T_{sn}$ (~120 K) of our FeSe film, newly prepared by pulsed laser deposition [37], also follow the universal linear relationship found in this work.

**Acknowledgments**

This work is supported by National Natural Science Foundation of China (projects 11574370 & 11190020), the National Basic Research Program of China (projects 2013CB921700 & 2016YFA0300301) and "Strategic Priority Research Program (B)" of the Chinese Academy of Sciences (No. XDB07020100).

**Author contributions**

X.L.D., F.Z. and Z.X.Z. planned and designed the project and summarized the data. D.N.Y., S.L.N. and Y.L.H. contributed to single crystal synthesis. Z.P.F. contributed to film growth. J.Y., D.N.Y. and Y.L.H. measured transport and magnetic data analyzed by K.J., J.Y. and X.L.D. with theoretical support from G.M.Z.. G.M.Z. introduced theoretical model. F.Z., X.L.D. and G.M.Z. wrote the paper. All authors provided comments for the paper.

**References**

[1] F. C. Hsu, J. Y. Luo, K. W. Yeh, T. K. Chen, T. W. Huang, P. M. Wu, Y. C. Lee, Y. L. Huang, Y. Y. Chu, D. C. Yan and M. K. Wu, Proc Natl Acad Sci U S A **105**, 14262 (2008).

[2] T. M. McQueen, Q. Huang, V. Ksenofontov, C. Felser, Q. Xu, H. Zandbergen, Y. S. Hor, J. Allred, A. J. Williams, D. Qu, J. Checkelsky, N. P. Ong and R. J. Cava, Phys. Rev. B **79**, 014522 (2009).

[3] S. Medvedev, T. M. McQueen, I. A. Troyan, T. Palasyuk, M. I. Eremets, R. J. Cava, S. Naghavi, F. Casper, V. Ksenofontov, G. Wortmann and C. Felser, Nat. Mater. **8**, 630 (2009).

[4] M. Bendele, A. Amato, K. Conder, M. Elender, H. Keller, H. H. Klauss, H. Luetkens, E. Pomjakushina, A. Raselli and R. Khasanov, Phys. Rev. Lett. **104**, 087003 (2010).

[5] J. P. Sun, K. Matsuura, G. Z. Ye, Y. Mizukami, M. Shimozawa, K. Matsubayashi, M. Yamashita, T. Watashige, S. Kasahara, Y. Matsuda, J.-Q. Yan, B. C. Sales, Y. Uwatoko, J.-G. Cheng and T. Shibauchi, arXiv1512.06951 (2015).

[6] T. Terashima, N. Kikugawa, S. Kasahara, T. Watashige, Y. Matsuda, T. Shibauchi and S. Uji, arXiv1603.03487 (2016).

[7] B. Lei, J. H. Cui, Z. J. Xiang, C. Shang, N. Z. Wang, G. J. Ye, X. G. Luo, T. Wu, Z. Sun and X. H. Chen, Phys. Rev. Lett. **116**, 077002 (2016).

[8] X. F. Lu, N. Z. Wang, H. Wu, Y. P. Wu, D. Zhao, X. Z. Zeng, X. G. Luo, T. Wu, W. Bao, G. H. Zhang, F. Q. Huang, Q. Z. Huang and X. H. Chen, Nat. Mater. **14**, 325 (2015).

[9] X. Dong, H. Zhou, H. Yang, J. Yuan, K. Jin, F. Zhou, D. Yuan, L. Wei, J. Li, X. Wang, G. Zhang and Z. Zhao, J. Am. Chem. Soc. **137**, 66 (2015).

[10] X. Dong, K. Jin, D. Yuan, H. Zhou, J. Yuan, Y. Huang, W. Hua, J. Sun, P. Zheng, W. Hu, Y. Mao, M. Ma, G. Zhang, F. Zhou and Z. Zhao, Phys. Rev. B **92**, 064515 (2015).

[11] A. E. Böhmer and C. Meingast, arXiv1505.05120 (2015).

[12] J. K. Glasbrenner, I. I. Mazin, H. O. Jeschke, P. J. Hirschfeld, R. M. Fernandes and R. Valenti, Nat. Phys. **11**, 953 (2015).

[13] A. V. Chubukov, R. M. Fernandes and J. Schmalian, Phys. Rev. B **91**, 201105 (2015).

[14] M. A. Tanatar, A. E. Bohmer, E. I. Timmons, M. Schutt, G. Drachuck, V. Taufour, S. L. Bud'ko, P. C. Canfield, R. M. Fernandes and R. Prozorov, arXiv1511.04757 (2015).

[15] Y. Hu, X. Ren, R. Zhang, H. Luo, S. Kasahara, T. Watashige, T. Shibauchi, P. Dai, Y. Zhang, Y. Matsuda and Y. Li, Phys. Rev. B **93**, 060504 (2016).

[16] A. V. Chubukov, M. Khodas and R. M. Fernandes, arXiv1602.05503 (2016).

[17] M. Abdel-Hafiez, Y. J. Pu, J. Brisbois, R. Peng, D. L. Feng, D. A. Chareev, A. V. Silhanek, C. Krellner, A. N. Vasiliev and X.-J. Chen, Phys. Rev. B **93**, 224508 (2016).

[18] T. Shimojima, Y. Suzuki, T. Sonobe, A. Nakamura, M. Sakano, J. Omachi, K. Yoshioka, M. Kuwata-Gonokami, K. Ono, H. Kumigashira, A. E. Böhmer, F. Hardy, T. Wolf, C. Meingast, H. v. Löhneysen, H. Ikeda and K. Ishizaka, Phys. Rev. B **90**, 121111 (2014).

[19] S. H. Baek, D. V. Efremov, J. M. Ok, J. S. Kim, J. van den Brink and B. Buchner, Nat. Mater. **14**, 210 (2015).

[20] M. C. Rahn, R. A. Ewings, S. J. Sedlmaier, S. J. Clarke and A. T. Boothroyd, Phys. Rev. B **91**, 180501 (2015).

[21] Q. Wang, Y. Shen, B. Pan, X. Zhang, K. Ikeuchi, K. Iida, A. D. Christianson, H. C. Walker, D. T. Adroja, M. Abdel-Hafiez, X. Chen, D. A. Chareev, A. N.Vasiliev and J. Zhao, Nat.

Commun. **7**, 12182 (2016).

[22] Q. Wang, Y. Shen, B. Pan, Y. Hao, M. Ma, F. Zhou, P. Steffens, K. Schmalzl, T. R. Forrest, M. Abdel-Hafiez, X. Chen, D. A. Chareev, A. N. Vasiliev, P. Bourges, Y. Sidis, H. Cao and J. Zhao, Nat. Mater. **15**, 159 (2016).

[23] D. Yuan, Y. Huang, S. Ni, H. Zhou, Y. Mao, W. Hu, J. Yuan, K. Jin, G. Zhang, X. Dong and F. Zhou, Chin. Phys. B **25**, 077404 (2016).

[24] M. W. Ma, D. N. Yuan, Y. Wu, X. L. Dong and F. Zhou, Physica C: Superconductivity and its Applications **506**, 154 (2014).

[25] S. Rößler, C. Koz, L. Jiao, U. K. Rößler, F. Steglich, U. Schwarz and S. Wirth, Phys. Rev. B **92**, 060505 (2015).

[26] C. Fang, H. Yao, W.-F. Tsai, J. Hu and S. A. Kivelson, Phys. Rev. B **77**, 224509 (2008).

[27] C. K. Xu and S. Sachdev, Nat. Phys. **4**, 898 (2008).

[28] F. Wang, S. A. Kivelson and D.-H. Lee, Nat. Phys. **11**, 959 (2015).

[29] R. Yu and Q. Si, Phys. Rev. Lett. **115**, 116401 (2015).

[30] Y. Sun, S. Pyon and T. Tamegai, Phys. Rev. B **93**, 104502 (2016).

[31] T. Urata, Y. Tanabe, K. K. Huynh, Y. Yamakawa, H. Kontani and K. Tanigaki, Phys. Rev. B **93**, 014507 (2016).

[32] T. M. McQueen, A. J. Williams, P. W. Stephens, J. Tao, Y. Zhu, V. Ksenofontov, F. Casper, C. Felser and R. J. Cava, Phys. Rev. Lett. **103**, 057002 (2009).

[33] A. E. Böhmer, F. Hardy, F. Eilers, D. Ernst, P. Adelmann, P. Schweiss, T. Wolf and C. Meingast, Phys. Rev. B **87**, 180505 (2013).

[34] C. Koz, M. Schmidt, H. Borrmann, U. Burkhardt, S. Rößler, W. Carrillo-Cabrera, W. Schnelle, U. Schwarz and Y. Grin, Z. Anorg. Allg. Chem. **640**, 1600 (2014).

[35] K. Kothapalli, A. E. Bohmer, W. T. Jayasekara, B. G. Ueland, P. Das, A. Sapkota, V. Taufour, Y. Xiao, E. E. Alp, S. L. Bud'ko, P. C. Canfield, A. Kreyssig and A. I. Goldman, arXiv1603.04135 (2016).

[36] U. Pachmayr, N. Fehn and D. Johrendt, Chem. Commun. **52**, 194 (2016).

[37] Z. Feng, J. Yuan and K. Jin, in preparation.

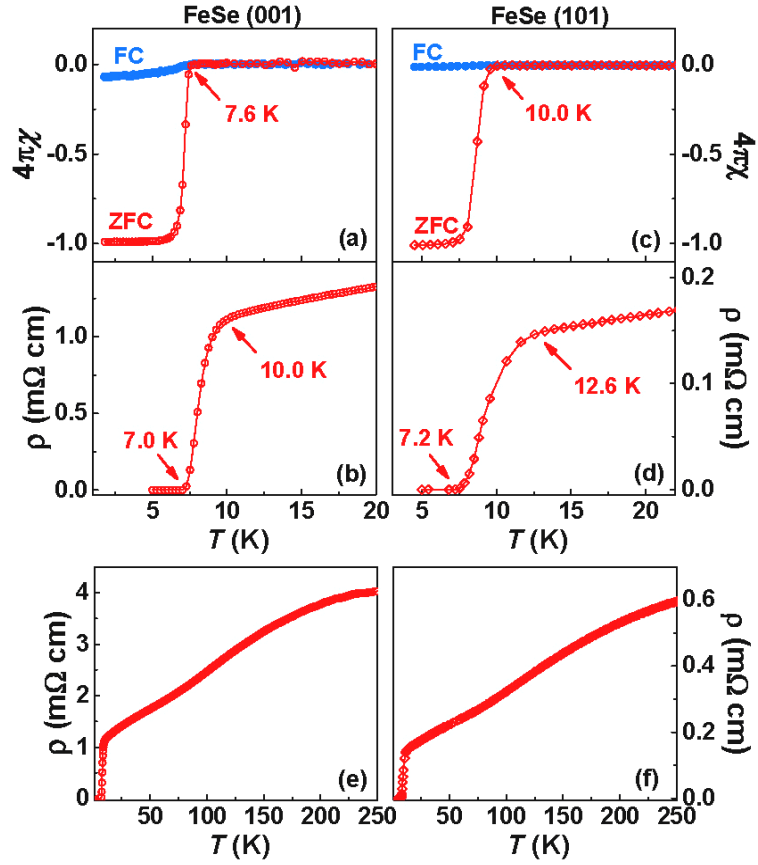

FIG. 1. Temperature dependences of the static magnetic susceptibility, corrected for demagnetization factor, and the electrical resistivity for the two typical FeSe single crystal samples with (001) and (101) orientations. (**a**) and (**c**) the data of magnetic susceptibility; (**b**) and (**d**) the electrical resistivity near the superconducting transitions; (**e**) and (**f**) the resistivity up to 250 K. The magnetic measurements are performed under $H = 1$ Oe.

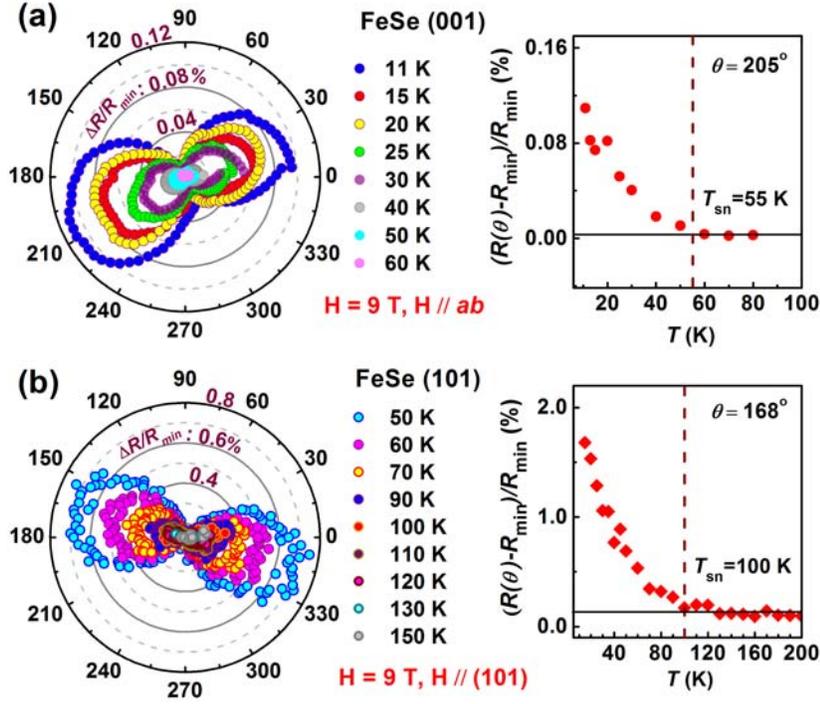

FIG. 2. Temperature dependences of the angular-dependent magnetoresistance showing the two-fold rotational symmetry below ~55 K (the upper right) and ~100 K (the lower right). (**a**) The FeSe crystal sample of (001) orientation with $T_c$ ~7.6 K. (**b**) the sample of (101) orientation with $T_c$ ~10.0 K. The $\theta$ is the angle between the directions of the external field (*H*) and the current (*I*), with $\theta = 0°$ corresponding to $H \perp I$.

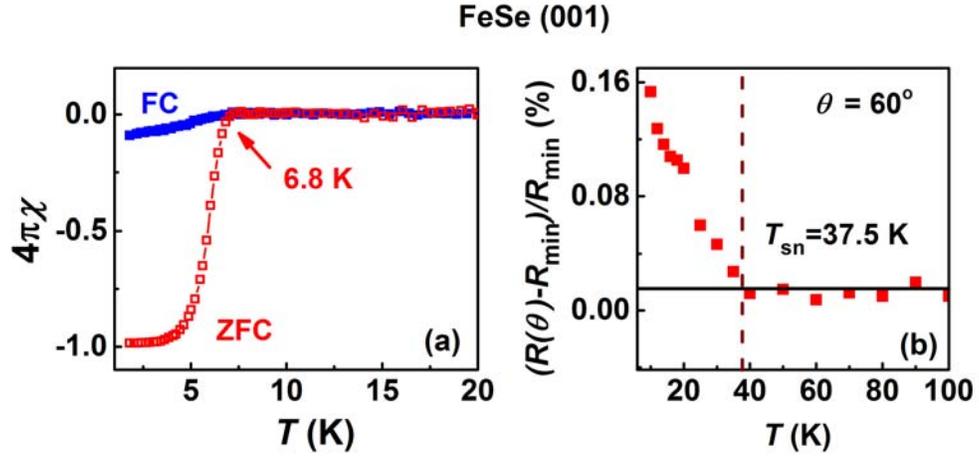

FIG. 3. (**a**) Temperature dependence of the static magnetic susceptibility for the hydrothermal FeSe crystal sample of (001) orientation with the $T_c \sim 6.8$ K. (**b**) Its temperature dependence of the maxima in anisotropic AMR showing the $T_{sn} \sim 37.5$ K. The $\theta$ is the angle between the directions of the external field (*H*) and the current (*I*), with $\theta = 0°$ corresponding to $H \perp I$.

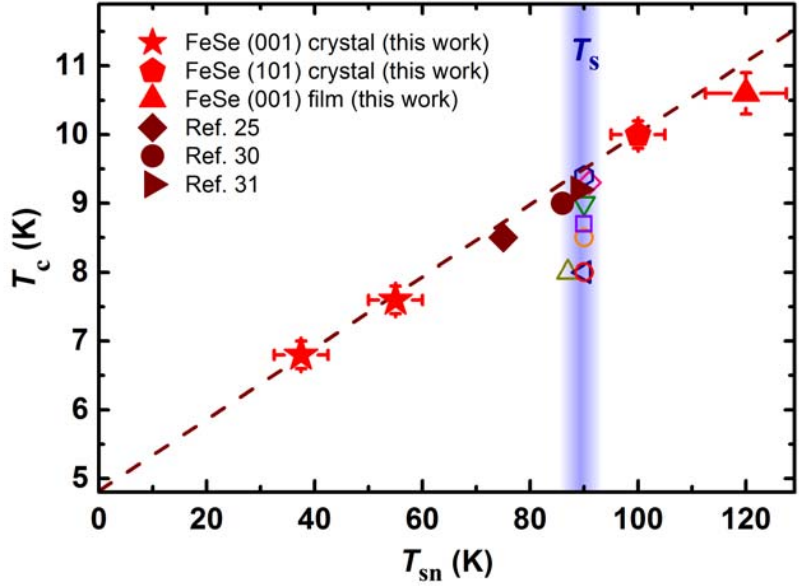

FIG. 4. The universal linear relationship between the superconducting transition temperature ($T_c$) and the Ising spin-nematic ordering temperature ($T_{sn}$) among various FeSe samples (the solid symbols). The hollow symbols in the vertical blue-shaded area represent the structure phase transition temperatures ($T_s$'s) by the x-ray or neutron diffractions on various FeSe samples of different $T_c$'s [19, 20, 22, 32-36]. The experimental uncertainty for the $T_c$ values of our crystal samples, defined as the temperatures where the shielding and Meissner signals clearly separate from one another, is estimated as ±0.25 K from the signal responses. The $T_{sn}$'s have an estimated error < ±5 K, which is the temperature sampling interval.